\documentstyle[amsfonts,pra,aps]{revtex}
\begin{document}
%\preprint{Volume XX, Number YY\hspace{.5in}PHYSICAL REVIEW LETTERS
%\hspace{1.05in}Day Month, 2000}
\twocolumn[\hsize\textwidth\columnwidth\hsize\csname@twocolumnfalse%
\endcsname
%\draft command makes pacs numbers print
\draft
\title{Ipso-information-transfer}
%\author{John V. Corbett*}
\author{John V. Corbett
\cite{byline1}}
%\footnote[1]{Email address:jvc@ics.mq.edu.au}}
\address{Department of Mathematics,Macquarie University,
Sydney, N.S.W. 2109, Australia}
%\author{Dipankar Home$^\dagger$}
\author{Dipankar Home
\cite{byline2}}
%\footnote[1]{Email address:dhom@boseinst.ernet.in}}
\address{Department of Physics, Bose Institute, Calcutta
700 009, India}
\date{\today}
\maketitle
\begin{abstract}
A tripartite system with entangled 
non-orthogonal states is used to transfer retrievable or usable 
information without requiring an external channel,{\em
ipso}-information-transfer(IIT). The non-Schmidt decomposable 
entanglement couples two independent interactions through 
which the information is transferred between a 
pair of non-interacting components of the system. 
We outline a dynamical model of IIT using localised
particles. Implications for quantum nonlocality are raised. 

\end{abstract}

]

% body of paper here
%\section{Introduction}

In recent years the use of quantum entanglement as a 
resource for transferring and processing information 
has attracted increasing attention. Much of which has
been on studies of quantum teleportation, QT, using 
discrete or continuous variables\cite{1}. In QT the 
information transfer from one wing to another 
involves  a maximally entangled state in conjunction 
with a Bellbasis measurement in one wing and an 
external  channel through which information about 
the outcome of this measurement is classically 
communicated to the other wing.  The information  
transfer is attained by transporting a quantum 
state in which the information is encoded. 

In this paper 
we show that the quantum formalism permits a 
{\em different} protocol involving a tripartite 
entanglement in which the information transfer  
does {\em not} require any external channel 
and does {\em not} 
involve the transfer of a quantum state from 
one system to another. In it, 
the Bellbasis measurement is replaced  by an 
interaction between two subsystems, 2 and 3, 
in one wing while 2 is entangled with subsystem 1 
in the other wing. The external channel is replaced 
by a  long range interaction between 1 and 2.  

In our scheme, called ipso-information transfer
(IIT), particles 1 and 2 are first prepared in an 
entangled state. They separate but continue to 
interact via a long-range residual interaction, 
$V_{R}(1,2)$. While 
this interaction continues, 
particle 1 flies off to Bob and a separate 
interaction, $V_{N}(2,3)$,
between 2 and another particle 3 may be switched 
on by Alice. {\em If} Alice switches the interaction on 
then the joint state of 1, 2 and 3 becomes an 
entangled tripartite state. Then, after all the 
particles have separated and have ceased interacting, 
1 and 3 are entangled with each other because both 
are in turn entangled with 2. The pairs of 
single particle states of 2 that are used 
are not orthogonal. 
Hence this tripartite entangled state is 
{\em not} Schmidt decomposable, and so it is not  
asymptotically equivalent to a GHZ state \cite{2}.

We note that in IIT there are two 
levels of information transferred from Alice to Bob:
1.The {\em yes/no} decision taken by Alice whether or not
to turn on the interaction, $V_{N}(2,3)$.
2. Contingent on a {\em yes} decision, information about 
the final states of 3 given by their inner product. 
Provided that Bob knows beforehand the form of the 
interaction $V_{N}(2,3)$, he can
determine its strength.   This 
information is received by Bob when he measures the 
expectation values of a certain class of 
observables of 1, provided that he knows the initial
entangled state of particles 1 and 2. The difference 
in the expectation values measured by Bob 
when Alice has or has not turned on the interaction
is proportional to the inner product of the final states 
of 3. For this protocol to work, as well
as the tripartite entanglement, both the 
dynamical interactions $V_{R}(1,2)$ and $V_{N}(2,3)$ 
are necessary. But its working does not require any 
extra communication between Alice and Bob, say by 
a phone call, as is required in QT.

The other significant feature of this example, which 
is discussed at the conclusion of the paper, is that 
it implies a form of quantum nonlocality which  
is quite different from those previously studied.

%\section{Communicating through an entangled tripartite state}

We begin by giving an outline of the IIT scheme. Initially 
the particles 1 and 2 are prepared in an entangled
state $\Psi(1,2)$,   

\begin{equation}
\Psi(1,2)=a\psi_{+}(1)\phi_{+}(2)
+b\psi_{-}(1)\phi_{-}(2),\label{1}
\end{equation}
where each of the single particle wave functions is 
normalised and 
$\langle\psi_{+}(1)|\psi_{-}(1)\rangle=0$,
however,  
$\langle\phi_{+}(2)|\phi_{-}(2)\rangle=\gamma_{2}\neq{0}$.
We take $|a|^{2}+|b|^{2}=1$ so that 
$\Psi_{0}(1,2)$ is also normalised.

Let the state $\Psi(1,2)$ be prepared by Alice prior to the
time $t_{1}$. For times $t>t_{1}$, 1 and 2 
become separated spatially but continue to interact 
through $V_{R}(1,2)$ whose strength diminishes as the
separation between 1 and 2 grows. 
While the interaction $V_{R}(1,2)$ is still 
non-negligible compared to the kinetic energy
 2 and 3 interact,
during the period from
$t_{2}$ to $t_{3}$, to form a
tripartite entanglement. At times $t>t_{3}$, 2 and 3 have
ceased interacting and are moving apart. At a later 
time $t_{4}$, when Bob receives 1, 
$V_{R}(1,2)$ is also negligible in comparison with the 
kinetic energy terms and all the particles 1, 2 and 3
are non-interacting.

The final state is the normalised
wave function,
\begin{equation}
\Psi_{f}(1,2,3)=a\psi_{+}(1)\phi_{+}(2)\chi_{+}(3)
+b\psi_{-}(1)\phi_{-}(2)\chi_{-}(3).\label{6}
\end{equation}
Each of the single particle wave functions has norm 1, but
only
$\langle\psi_{+}(1)|\psi_{-}(1)\rangle=0$, whereas,
$\langle\phi_{+}(2)|\phi_{-}(2)\rangle=\gamma_{2}\neq{0}$
and
$\langle\chi_{+}(3)|\chi_{-}(3)\rangle=\gamma_{3}$ may
be non-zero.

At a time $t>t_{4}$, Bob measures an observable of 1,
represented by a self-adjoint operator
$A(1)$ on ${\cal H}(1)$,which satisfies
$\langle\psi_{+}(1)|A(1)|\psi_{-}(1)\rangle=\alpha\neq{0}$.
The expectation value of $A(1)$ in the final state
$\Psi_{f}(1,2,3)$ is
\begin{eqnarray}
\langle\Psi_{f}(1,2,3)|A(1)|\Psi_{f}(1,2,3)\rangle=
|a|^{2}\langle\psi_{+}(1)|A(1)|\psi_{+}(1)\rangle
\nonumber\\ +
|b|^{2}\langle\psi_{-}(1)|A(1)|\psi_{-}(1)\rangle +
2\Re{e}\ \overline{a}b\gamma_{2}\gamma_{3}\alpha.\label{7}
\end{eqnarray}

This result is to be contrasted with that which arises 
if particles 2 and 3 had {\em not} interacted. In that 
case the tripartite system would have a final state 
$\Psi_{f}^{'}(1,2,3)=\Psi(1,2)\chi_{0}(3)$
where $\chi_{0}(3)$ is normalised and
$\Psi(1,2)$ is given by equation (\ref{1}).

The expectation value of $A(1)$ in $\Psi_{f}^{'}(1,2,3)$ is 
\begin{eqnarray}
\langle\Psi_{f}^{'}(1,2,3)|A(1)|\Psi_{f}^{'}(1,2,3)\rangle=
|a|^{2}\langle\psi_{+}(1)|A(1)|\psi_{+}(1)\rangle
\nonumber\\ +
|b|^{2}\langle\psi_{-}(1)|A(1)|\psi_{-}(1)\rangle +
2\Re{e}\ \overline{a}b\gamma_{2}\alpha.\label{8}
\end{eqnarray}
The difference $\delta$ between the values in 
(\ref{7}) and (\ref{8}) is,
\begin{equation}
\delta=2\Re{e}\
\overline{a}b\gamma_{2}(1-                        
 \ \overline\gamma_{3})\alpha.\label{9}
\end{equation}
 Once the experimenter knows $\alpha$ and $\Psi(1,2)$, 
ie a, b and $\gamma_{2}$, then $\gamma_{3}$ can be
determined by measuring $\delta$.

In order to illustrate the way the information is 
transferred from Alice to Bob, we consider 
the following scenario. Alice
has prepared an entangled state, 
$\Psi(1,2)$, of 1 and 2 and is about to send 
1 off to a distant observer Bob. 
Before doing this, she  
chooses whether to switch on an interaction, $V_{N}(2,3)$, 
between 2 and another particle 3, while $V_{R}(1,2)$, 
between 1 and 2, remains at a non-negligible strength
compared to the kinetic energy terms. 

Bob can determine which choice Alice had made by measuring
the expectation value of $A(1)$. The expectation value
depends upon $\gamma(3)$, by equation (\ref{7}), 
if  2 and 3 interact while 1 and 2 are still interacting  
but is independent of $\gamma(3)$, by equation (\ref{8}),
if 2 and 3 do not. 

The information about whether Alice switched on 
$V_{N}(2,3)$ and, if she had, about the value of $\gamma(3)$ 
is transferred to Bob without 1 ever interacting with 3 and 
after 1 has ceased interacting with 2. 
Alice and Bob need no communication 
channel external to the tripartite 
system.

%\section{The preparation of the entangled states}

Now, we proceed to describe an explicit dynamical model
of the IIT scheme. For this purpose we use time-dependent 
coordinate wave 
functions because the interaction $V_{R}(2,3)$ whose 
strength depends upon the {\em spatial separation} 
is crucial to our scheme. The position coordinate space 
is one
dimensional to simplify the notation.

 The initial state of
particle 1 is 
\begin{equation}
\zeta(q(1),t)=a\psi_{+}(q(1),t)
+b\psi_{-}(q(1),t),\label{10}
\end{equation}
where both $\psi_{+}(q(1),t)$ and
$\psi_{-}(q(1),t)$ are normalised and are orthogonal 
to each other so that
$\zeta(q(1),t)$ is normalised when $|a|^{2}+|b|^{2}=1$.
We assume that the wave function $\psi_{+}(q(1),t)$ 
has support in the open interval $I_{+}$ while
$\psi_{-}(q(1),t)$ has support in the open interval 
$I_{-}$ with a 
non-zero separation $d$ between the intervals. 
The particle 2 has the initial wave function
$\phi_{0}(q(2),t)$.

The interaction between 1 and 2, that 
produces $\Psi(1,2)$ of equation (\ref{1}), is turned 
on at time $t_{0}$. We use  
the von Neumann interaction $V_{N}(1,2)$,
\begin{equation}
V_{N}(1,2)=g(1,2)(q(1){\cdot}p(2)).\label{11}
\end{equation}
 von Neumann used this type of interaction to describe a
measurement process\cite{3}. $g(1,2)$ is the coupling
constant, $q(1)$ is the position operator for particle 1, 
and $p(2)$ is the momentum operator for particle 2. 1 
and 2 also interact through a weaker, long range 
potential $V_{R}(1,2)$. The unitary operator
$U(1,2)$ governing the  evolution of the particles is 
generated by the Hamiltonian operator $H$,
\begin{equation}
H = {p^{2}(1)\over{2m_{1}}} +{p^{2}(2)\over{2m_{2}}}
+V_{R}(1,2) +V_{N}(1,2).\label{12}
\end{equation}

We assume that $g(1,2)$ is so large  
that both $V_{R}(1,2)$ and the kinetic energy terms in
$H$ can be neglected in comparison to
$V_{N}(1,2)$. Moreover, it will be seen that 
$V_{N}(1,2)$ is impulsive. 
While it acts the total Hamiltonian
is effectively 
\begin{equation}
H = V_{N}(1,2).\label{13}
\end{equation}

If $V_{N}(1,2)$ is switched on from $t_{0}$ to $t_{1}$, 
the relevant Schroedinger equation is 
\begin{equation}
{\hbar\over\imath}{\partial\Psi\over\partial t}(q(1),q(2),t)
=-{\hbar\over\imath}g(1,2)q(1){\partial\Psi\over\partial q(2)}
(q(1),q(2),t).\label{14} 
\end{equation}
This simplifies to a transport equation\cite{4} whose 
general solution, for $t_{0}\leq{t}\leq{t}_{1}$, is 
well known to be,
\begin{equation}
\Psi(q(1),q(2),t)=f(q(1),q(2)-g(1,2)(t-t_{0})q(1))\label{15a}
\end{equation}
where $f$ is a function of two variables.
Given the initial wave function 
\begin{eqnarray}
\Psi(q(1),q(2),t_{0})=a\psi_{+}(q(1),t_{0})\phi_{0}(q(2),t_{0})
\nonumber\\
+b\psi_{-}(q(1),t_{0})\phi_{0}(q(2),t_{0}),\label{15b}
\end{eqnarray}
the solution of the initial value problem for the
Schrodinger equation (\ref{14}) is,  
\begin{eqnarray}
\Psi(q(1),q(2),t)=a\psi_{+}(q(1),t_{0})\phi_{0}(q(2)-d(1,2)
q(1),t_{0})\nonumber\\
+b\psi_{-}(q(1),t_{0})\phi_{0}(q(2)-d(1,2)q(1),t_{0}),
\label{16}
\end{eqnarray}
for $t_{0}\leq{t}\leq{t_{1}}$ and $d(1,2)=g(1,2)(t_1-t_0)$.
This can be written in the form (\ref{1}) of an entangled 
bipartite state,
\begin{eqnarray}
\Psi(q(1),q(2),t_{1})=a\psi_{+}(q(1),t_{0})\phi_{+}(q(2),t_{0})\nonumber\\
+b\psi_{-}(q(1),t_{0})\phi_{-}(q(2),t_{0})\label{17}
\end{eqnarray}
where the wave functions $\phi_{s}(q(2),t_{0})$, for $s=\pm$, are given by
weighted shifts of the  argument $q(2)$ of $\phi_{0}(q(2),t_{0})$,
\begin{equation}
\phi_{s}(q(2),t_{0})=
\int|\psi_{s}(q(1),t_{0})|^{2}\phi_{0}(q(2)-d(1,2)
q(1),t_{0})dq(1).\label{19a}
\end{equation}
The formulae are correct because the supports of $\psi_{+}(q(1),t_{0})$
and $\psi_{-}(q(1),t_{0})$ are disjoint. Moreover,
because  $\phi_{0}(q(2),t_{0})$ and $\Psi(q(1),q(2),t_{1})$ 
are normalised so are 
$\phi_{+}(q(2),t_{0})$ and $\phi_{-}(q(2),t_{0})$. A proof
of their normalisation uses Young's inequality\cite{5} for the 
$L_{2}$ norm of the convolution of an $L_{1}$ function with an 
$L_{2}$ function and the fact that  $|a|^{2}+|b|^{2}=1$. This 
completes the preparation of the bipartite state (\ref{1}).
$V_{N}(1,2)$ acted instantaneously 
because all wave functions on the left  of
(\ref{17}) have remained at the initial time $t_{0}$.

At the time $t_{2}>t_{1}$, 2 begins to interact with 3 while
 1 and 2 are still interacting through $V_{R}(1,2)$.
The interaction between 2 and 3 is taken to be 
another von Neumann interaction $V_{N}(2,3)$.   
The Hamiltonian for the three particle system is ,
\begin{equation}
H = {p^{2}(1)\over{2m_{1}}} + {p^{2}(2)\over{2m_{2}}}
+ {p^{2}(3)\over{2m_{3}}} +  V_{N}(2,3) +
V_{R}(1,2).\label{20}
\end{equation}
The operator $p(j)$ is the momentum operator of the $j$th
particle. $V_{N}(2,3)$ is given by 
\begin{equation}
V_{N}(2,3)=g(2,3)(q(2){\cdot}p(3)).\label{21}
\end{equation}
The coupling constant $g(2,3)$ is assumed
to be so large that 
both the kinetic energy terms and $V_{R}(1,2)$ are 
negligible in comparison to $V_{N}(2,3)$. The total 
Hamiltonian is effectively 
\begin{equation}
H = V_{N}(2,3).\label{22}
\end{equation}
If  $V_{N}(2,3)$ is switched on from $t_{2}$ to $t_{3}$, 
the relevant Schroedinger equation 
\begin{equation}
{\hbar\over\imath}{\partial\Psi\over\partial t}
=-{\hbar\over\imath}g(2,3)q(2){\partial\Psi\over\partial q(3)}
(q(1),q(2),q(3),t).\label{23} 
\end{equation}
simplifies to a transport equation whose general
solution, for $t_{2}\leq t\leq t_{3}$, is\cite{7},
\begin{equation}
\Psi=
f(q(1),q(2),q(3)-g(2,3)(t-t_{2})q(2))
\label{24}
\end{equation}
where $f$ is a function of three variables.
Given that the initial wave function, 
$\Psi(q(1),q(2),q(3),t_{2})$, is
\begin{eqnarray}
a\psi_{+}(q(1),t_{2})\phi_{+}(q(2),
t_{2})\chi_{0}(q(3),t_{2})\nonumber\\
+b\psi_{-}(q(1),t_{2})\phi_{-}(q(2),t_{2})\chi_{0}(q(3),t_{2})
\label{24}
\end{eqnarray}
the solution, $\Psi(q(1),q(2),q(3),t)$,of the 
initial value problem for
equation (\ref{23}), for
$t_{2}\leq t\leq t_{3}$,is 
\begin{eqnarray}
a\psi_{+}(q(1),t_{2})\phi_{+}(q(2),t_{2})
\chi_{0}(q(3)-g(2,3)(t-t_{2})q(2),t_{2})\nonumber\\
+b\psi_{-}(q(1),t_{2})\phi_{-}(q(2),t_{2})\chi_{0}(q(3)-g(2,3)
(t-t_{2})q(2),t_{2})\label{25}
\end{eqnarray}
At $t=t_{3}$, the wave function,$\Psi(q(1),q(2),q(3),t_{3})$, 
is in the form (\ref{6}) of the final tripartite state, 
\begin{eqnarray}
a\psi_{+}(q(1),t_{2})\phi_{+}(q(2),
t_{2})\chi_{+}(q(3),t_{2})\nonumber\\
+b\psi_{-}(q(1),t_{2})\phi_{-}(q(2),t_{2})\chi_{-}(q(3),t_{2}).
\label{26}
\end{eqnarray}   
in which the wave functions $\chi_{s}(q(3),t_{2})$, for $s=\pm$,
are given by
weighted shifts of the  argument $q(3)$ of $\chi_{0}(q(3),t_{2})$.
\begin{eqnarray}
\chi_{s}(q(3),t_{2})=
\int|\psi_{s}(q(1),t_{2})|^{2}|
\phi_{s}(q(2),t_{2})|^{2}\times\nonumber\\
\chi_{0}(q(3)-g(2,3)(t_{3}-t_{2})q(2),t_{2})dq(1)dq(2).
\label{28}
\end{eqnarray}
The proof of (\ref{28}) is similar to that used to justify equation
(\ref{17}). Firstly, the product wave functions,
$\psi_{+}(q(1),t_{2})\phi_{+}(q(2),t_{2})$ and
$\psi_{-}(q(1),t_{2})\phi_{-}(q(2),t_{2})$, are
orthogonal at time $t_{2}$ because, by construction,
the product wave functions,
$\psi_{+}(q(1),t_{0})\phi_{+}(q(2),t_{0})$ and
$\psi_{-}(q(1),t_{0})\phi_{-}(q(2),t_{0})$, are
orthogonal at time $t_{0}$ and
the time evolution 
operator $U_{t}(1,2)$ is unitary.Again, the 
normalisation of $\chi_{+}(q(3),t_{2})$ and
$\chi_{-}(q(3),t_{2})$ follows from that of 
both $\chi_{0}(q(3),t_{2})$ and
$\Psi(q(1),q(2),q(3),t_{3})$ using Young's
inequality\cite{5}. $V_{N}(2,3)$ acted
instantaneously as all wave functions in 
(\ref{26}) stayed at time $t_{2}$.

After $t_{3}$, when the impulsive interaction $V_{N}(2,3)$ 
ceases acting, the tripartite wave function evolves
under the action of the Hamiltonian $H$,
\begin{equation}
H = {p^{2}(1)\over{2m_{1}}} + {p^{2}(2)\over{2m_{2}}}
+ {p^{2}(3)\over{2m_{3}}} + V_{R}(1,2)\label{29} 
\end{equation}
until $t_{4}$ when $V_{R}(1,2)$ becomes negligible in
comparison to the
kinetic energy terms in H. At any time $t>t_{4}$
Bob can measure the observable $A(1)$.  

%\section{Summary and Conclusions}

It is crucial for the transference of the 
information that 1 and 2 continued to interact 
while 2 and 3 were interacting. If  $V_{R}(1,2)$ 
had become negligible compared to the kinetic energy terms 
before $V_{N}(2,3)$ commenced operating, the final 
entangled state, $\Psi_{f}(1,2,3)$, given by (\ref{6}), 
could not be formed by a unitary dynamics because the 
states of 2 are not orthogonal \cite{6}. On the other 
hand,if the states of particle 2 are 
orthogonal,$\gamma_{2}=0$, and there is 
no transfer of usable information, as shown by  
(\ref{7}).

Furthermore, as well as this, two 
{\em independent} interactions 
$V_{R}(1,2)$ and  $V_{N}(2,3)$ are necessary for the 
above transfer.  They become linked via the 
entanglement. The magnitude of
$V_{R}(1,2)$ is {\em irrelevant} as long as it is 
not negligible compared
to the kinetic energy terms. 

The IIT scheme is robust in that 
parameters of the processes can be varied and 
information is still tranferred.  
For example, by changing 
the values of $g(1,2)$ and $g(2,3)$ the values of
$\gamma_{2}$ and $\gamma_{3}$
may be varied and the scheme still works. 
Transfer occurs unless $\gamma_{2}=0$.

Furthermore, the two levels of transferred information 
in the IIT scheme exhibit different responses to 
variations of the parameters. The first 
level is {\em not} sensitive to variations provided 
that $\gamma_{2}$ is non-zero but the second level is. 
The dependence of $\gamma_{3}$ on $g(2,3)$ can be  
calculated when $T$ is the duration of the interaction, 
$\phi_{+}(2)$ and $\phi_{-}(2)$ are Gaussians with means
$m_{+}$ and $m_{-}$ and the same variance $\sigma^{2}$, 
while $\chi_{0}(3)$ is a Gaussian with mean 0 and variance 
$\beta^{2}$. If $G=g(2,3)^{2}\times{T^{2}}/2$,then

\begin{equation}
\gamma_{3} = \beta^{-1}\times{\exp{-MG}}\label{29},
\end{equation}

where $M=(m_{+}-m_{-})^{2}/{2K(G)}$,
and $K(G)=(\beta^{2}+G\sigma^{2})$.
Thus even if Bob
knows neither the initial state of 3 nor the form of 
the interaction between  2 and 3, he can obtain some 
knowledge of the final states of  3, viz. the value 
of their inner product, using equation(\ref{7}).  But, 
if he knows both the initial state and the interaction 
he can measure the
product of the  strength and duration of the interaction 
using an equation like (\ref{29}).

In conclusion, we note that the 
above scheme embodies an unusual form of quantum 
nonlocality as shown by  
the following scenario. After the entangled state 
(\ref{1}) is prepared,  
1 and 2 go separately to Bob and Alice respectively. 
As well, Alice receives particle 3 from a third person, 
Carol, who had prepared 3 in a certain initial state, say, 
a Gaussian with mean 0 and variance $\beta$ as for 
equation (\ref{29}). Then Alice 
switches on the interaction $V_{N}(2,3)$. Subsequently, 
Bob measures an observable $A(1)$ whose expectation 
value is given by equation (\ref{7}). If Carol varied 
the preparation of the initial state of 3 (e.g., by 
varying the parameter $\beta$) but Alice kept the form 
and strength of $V_{N}(2,3)$ unchanged, then equations 
(\ref{7}) and (\ref{29}) imply that the  
value measured by Bob will change, while
Alice by her local measurements will not be able to 
detect any change.  

Thus a local change of parameter performed by Carol 
at a region spatially distant from Bob affects the 
statistical result of a measurement performed by Bob, 
even though the particles 1 and 3 never interact at 
any stage. Such an effect is nonlocal in the sense 
that it contradicts Einstein's locality condition, 
viz. "..... the real factual situation of the system 
S2 is independent of what is done with the system S1 
which is spatially separated from the former" \cite{7}.
According to any local model, the results of local 
measurement on particle 1 are determined entirely by 
the initial entangled state (\ref{1}) of 1,2 and the 
interaction $V_{R}(1,2)$ between them. Moreover, 
for a given interaction $V_{N}(2,3)$, the
final states of 2 in  
(\ref{6}) are unaffected 
by changes in the initial state of 3. Hence, in a
local model, such changes in 3  are not expected to
produce an observable change in the properties of 1.
This expectation is not satisfied in the IIT example.
The non-locality is independent  
of how "realism" is specified or what form of "hidden 
variables" is used in a local model.   

It is significant that, in IIT, nonlocality is discernible 
through statistical measurements in one of the wings 
without needing to use correlations between observations 
in the different wings. This contrasts with the usual 
arguments for quantum nonlocality (whether or not a 
Bell-type inequality is used)\cite{8}. 
Long range interactions in conjunction 
with kinematic entanglement produce this type of 
non-locality in non-Schmidt decomposable
states.  

%\section{Acknowledgements}
The authors acknowledge the financial support for D.H. via a
Research Grant to J.V.C. from Macquarie University where this
collaboration was initiated.  The research of D.H. is supported
by the Dept. of Science and Technology, Govt. of India.
Thanks are due to Sougato Bose of Claredon Laboratory, Oxford and 
also to Dale Woodside of Macquarie University for their helpful 
suggestions at different stages of the writing of this paper.

\end{document}